\documentclass[secnumarabic,amssymb, nobibnotes, aps, prd]{revtex4}%
\usepackage{graphicx}
\usepackage{dcolumn}
\usepackage{bm}
\usepackage{amsmath}%
\usepackage{amsfonts}%
\usepackage{amssymb}

\begin{document}

\title{On the ultra high energy cosmic rays and the origin of the cosmic microwave background radiation}

\author{C. E. Navia}
\address{Instituto de F\'{\i}%
sica Universidade Federal Fluminense, 24210-346,
Niter\'{o}i, RJ, Brazil}

\author{C. R. A. Augusto}
\address{Instituto de F\'{\i}%
sica Universidade Federal Fluminense, 24210-346,
Niter\'{o}i, RJ, Brazil}

\author{K. H. Tsui}
\address{Instituto de F\'{\i}%
sica Universidade Federal Fluminense, 24210-346,
Niter\'{o}i, RJ, Brazil}

\date{\today}

\begin{abstract}
Some inconsistencies to the assumption of a cosmological origin of the cosmic microwave background CMB,  such as the absence of gravitational lensing in the WMAP data, open the doors to some speculations such as a local origin to the CMB. We argue here that this assumption agrees with the absence of the GZK cutoff (at least  according to AGASA data) in the energy spectrum of the cosmic ray due to the cosmic interaction with the CMB at $6\times 10^{19} eV$ or above. Within 50 Mpc from Earth, the matter and light distributions are close to an anisotropic distribution, where the local cluster and local super-clusters of galaxies can be identified. In contrast, the ultra high energy comic rays data is consistent to an almost isotropic distribution, and there is no correlation between their arrival direction and astronomical sources within our local cluster. This means that the events above the GZK cutoff come from distances above $50$ Mpc, without an apparent energy loss. This scenario is plausible under the assumption of the CMB concentrated only within $3-4$ Mpc from Earth. In other words, the CMB has a local origin linked only to the local super-cluster of galaxies. In addition, the galactic and extragalactic energy spectra index within the energy equipartition theorem strongly constrains the dark matter and dark energy hypothesis, essential in the Big Bang cosmology. 
\end{abstract}


\maketitle

\section{Introduction}

After the discovery in 1965 of the cosmic microwave background radiation (CMB) and its interpretation as an afterglow of hot gases that filled the fledgling universe following the big bang and that fills the Universe as a sea of photons, with a mean temperature of $2.7^0$ K at present epoch, Greisen \cite{greisen66} and Zatsepin$\&$Kuz'min 
\cite{zapsepin66} independently pointed 
out that the CMB make the Universe opaque to ultra high energy cosmic rays (UHECR) particles with energies above $\sim 5 \times 10^{19}$ eV. 
Cosmic rays (i.e. protons) with energy above this energy are unable to propagate through
the CMB photons for a distance above 50Mpc due to the energy loss process, the pion production via the $\Delta$ resonance channel production
\begin{equation}
p+\gamma_{CMB}\rightarrow \Delta \rightarrow N \pi,
\end{equation}
where $N$ is a generic nucleon, the above process has a proton threshold energy of $\sim 6\times 10^{19}$ eV. There is also other important energy loss process such as $p+\gamma_{CMB}\rightarrow e^+e^-$ with a proton threshold energy of $5 \times 10^{17}$ eV. However, the energy loss in the last process (pair production) is significantly smaller than the energy loss by pion production. It was thus predicted that cosmic rays should cut-off at this GZK energy ($6\times 10^{19}$ eV), as well as, the energy spectrum of the UHECR will suffer modification by the interactions with the CMB.

But several events (air showers) originated by cosmic rays with primary energy above GZK cutoff
have been observed basically in all extensive air shower experiments, and the Auger project is expected to find hundreds more. No acceleration mechanism is known to such high energies within our local galaxy cluster and
they cannot be cosmological because of this GKZ cut-off. These characteristics have taken to formulation of new scenarios for the origin of the UHECR such as the violation of the Lorentz invariance \cite{coleman99} as responsible 
for the propagation of ultra high energy cosmic ray having energies above the GZK cutoff. In addition to the top-down models \cite{hill83}, the collapse or decay of super-massive particles, like magnetic monopoles, superconducting strings, as well as the $\nu$ Z burst \cite{fargion99} inside of a volume with a radius $\leq$ 50 MPc around the Earth can explain the UHECR data above the GZK cutoff. However, top-down models constrained by gamma ray flux, muon flux measurements favor primaries like hadrons (i.e protons).

The hypothesis that Gamma Ray Burst (GRB) might be responsible for the origin of UHECR has been suggest earlier \cite{waxman95}. The GRBs are probably the most powerfully events in the universe, and it is believe that protons can be accelerate in a GRB by internal shocks taking place in a collimated jet direction. So far, in short-GRBs, no afterglow and no redshift have been detected, suggesting that the short GRB sources are inside or close to our Galaxy. Consequently, a local source for a fraction of UHECR linked with these GRBs has been suggested \cite{navia05}. 
 
\section{Which is the origin of the cosmic microwave background radiation?}

The discovery of a pervasive background radiation in space by Penzias and Wilson \cite{penzias65} in 1965 is considered as the strongest evidence for the hot Big Bang model. The Cosmic Microwave Background Radiation (CMB) is a $2.7$ kelvin thermal black body spectrum with a peak in the micro wave range. 
The CMB is considered a relic of the Big Bang, whose origin is linked to the so called ``decoupling era''. 
In the past when the Universe was much smaller, the radiation was also much hotter. As the Universe expanded, it cooled down until it is today. 
In Penzias-Wilson's data the radiation appeared as highly isotropic.
However, in the next round of experiments \cite{conklin69,henry71,smoot77} temperature anisotropies were found. These anisotropies are expressed using the spherical harmonic expansion. The higher anisotropy is like dipole component
with $\Delta T/T \sim 10^{-3}$, and is interpreted as due to the motion of the Earth with respect to the 
``CMB rest frame'' with a velocity of $370\;km/s$, but the possibility (at least in part) for an intrinsic variation of the CMB radiation itself can not be eliminated. Excluding this component, the CMB is isotropic with small fluctuations
up to $\Delta T/T \sim 10^{-5}$. This result is interpreted as the strongest evidence in support of the cosmological principle, the basic assumption of cosmology that the universe is isotropic and homogeneous on a large scale.

However, after the COBE and specially the WMAP data \cite{spergel03} were available to the scientific community, some inconsistencies between the CMB data and its claimed cosmological origin have been pointed out by several authors. The main ones are indicated below:
\begin{itemize} 
\item $No\;lensing\;effect\;on\;the\;CMB$.
When light or any radiation from a distant galaxy is bent by gravity as it passes another galaxy or galaxy cluster, these distortions can appear as Einstein rings or week lensing shear effects. On the other hand, 
the most accepted interpretation for the cold spots observed in the WMAP's data on the CMB is that these cold spots are the birthmarks of galaxies and clusters of galaxies that condensed out of the primordial plasma. A large portion of the mass in the nearby universe is concentrated in small volumes of space. These are galaxies and massive galaxy clusters, which are surrounded by vast empty voids of intergalactic space. Consequently, radiation from some cold spots
would travel through mostly empty space and would look small by the time that radiation reached Earth. However, radiation from other cold spots, wold pass around or near massive gravity lenses. These focused spots would appear to be larger than the average spot. The problem is that this dispersion of sizes is not seen in the WMAP data. The cold
spots appear to be no lensing effect whatsoever.

Lieu and Mittaz \cite{lieu05} have calculated the all-sky variation in size of the CMB acoustic peaks based upon the lensing effect of clusters of galaxies, and compared with the WMAP data. They have concluded that a rather large lensing induced dispersion in the angular size of the primary acoustic peaks of the CMB power spectrum is inconsistent with WMAP observations, or in other words, a gravitational lensing effect is absent in the cosmic background radiation.

\item $The\;cluster\;shadow\;effect\;on\;the\;CMB$.
 A survey on the Sunyaev-Zel'dovich effect on 32 clusters of galaxies made by Lieu et al. \cite{lieu05b} have shown that the CMB from behind the clusters is slightly shadowed (only one quarter of that predicted) by hot electrons in the cluster. They have concluded that there were no strong evidences for an emission origin of the CMB at locations beyond the average redshift
 of our cluster sample (i.e. $z\sim 0.1$).

\item $The\;CMB\;alignments$.
Some analysis \cite{lerner95} on the CMB behavior have shown various unexpected alignments of the planes of the quadrupole and octupole moments with each other and with the direction of the dipole, as well as, with the alignment of the local super-cluster. This behavior in the CMB strongly constrained the assumption that the CMB is of cosmological origin. The alignments of the CMB strongly support a local origin of the CMB, linked with the local super-cluster.

This  subject is also comment by  Wibig et al. \cite{wibig05} analyzing the power spectrum of WMAP data for the two galactic hemispheres, North and South and to allow observation of the fact that there is a Southern excess at high latitudes. They conclude that the CMB asymmetries change strongly the canonical cosmological parameters.

\item $The\;CMB\;temperature\;in\;the\;COBE\;experiment$. The CMB temperature, from the dipole measurement, is significantly lower than the value obtained from the monopole \cite{robitaille07}. If all statistically significance is not removed in the COBE data, the CMB temperatures, derived from the monopole and from the dipole, are irreconcilable. This difference arises from a probable contamination of the CMB monopole component with the microwave signal emanating from the Earth oceans.  In short, all these evidences constitute a challenge to the interpretation of a cosmological origin for the CMB.

\item $The\;Big\;Bang\;Machine$. The Relativistic Heavy Ion Collider (RHIC) at Brookhaven National Laboratory, where the main objective is to re-creates the immediate aftermath of the Big Bang, the so called quark gluon plasma. So far, there are only reports of an obscure signature of transition to quark gluon plasma \cite{arsene05}. It is claimed that still there are experimental obstacles making it hard to directly observe the phase transition. Even so, the RHIC results have shown that, if the quark gluon plasma phase exist, it is not a gas. It behaves like a nearly ideal liquid fluid, in contrast to all predictions of the Big Bang cosmology on the basis of a gaseous quark gluon plasma.

\end{itemize}

\section{The cosmic ray energy spectrum and the energy equipartition theorem}

The main results of the CMB on the basis of WMAP data are the cosmological parameters of Big Bang cosmology picture. According to this scheme only 4.4\% of the matter in the Universe is in the form of baryons (ordinary matter), and  27\% corresponds to an exotic form of matter called as dark matter, and 73\% corresponds to the so called dark energy. In spite of 27 years of search for direct evidences specially of this exotic form of matter, so far, the results are negative \cite{akerib04}, and only upper limits have been obtained. 

We present here an analysis about the distribution of energy in the galaxy and which can be extended under certain circumstances to the whole Universe, on the basis of the cosmic ray energy spectrum. The results strongly constrain the dark matter and dark energy hypothesis.  
Under the assumption that the Galaxy is out of the formation phase and if $V$ denote the Galaxy's volume and W its energy contained in its volume. The rate of energy change is just the radiation rate J
\begin{equation}
\frac{dW}{dt}=-J.
\end{equation}  
The energy W is in the form of several modes, such as the turbulent, electromagnetic, cosmic rays, and even other exotic modes like dark matter and dark energy. However, a substantial  fraction of dark matter in the galaxy as is claimed for taking into account the galaxy rotation curve, disagree with the index of the galactic cosmic ray energy spectrum.   
This conclusion comes from the energy Equipartition theorem  that supplies the relation 
\begin{equation}
W=n\times U_n,
\end{equation}
where $n$ is the number of energy modes  and $U_n$ represents the nth mode of energy. Already in the sixties Syrovatsky \cite{syrovatsky61} had postulated three energy modes, called as the non thermal modes such as
turbulent, electromagnetic and cosmic rays. Consequently, the last equation can be written as
\begin{equation}
W=3U\equiv 3\bar{E}N,
\end{equation}
where $\bar{E}$ is the average energy of cosmic ray and
\begin{equation}
\frac{dW}{dt}=3\left(\frac{d\bar{E}}{dt}N+\bar{E}\frac{dN}{dt}\right),
\end{equation}
since the rate of energy variation can be written also as
\begin{equation}
\frac{dW}{dt}=-J\equiv \bar{E}\frac{dN}{dt},
\end{equation}
these last two equations give us the relation 
\begin{equation}
\frac{dN}{N}=-\frac{3}{2}\frac{d\bar{E}}{\bar{E}},
\end{equation}
and after integration we have
\begin{equation}
N=N_0 E^{-3/2}.
\end{equation}
This index $3/2(=1.5)$ is not far of the experimental index of the integral power energy spectrum of galactic cosmic ray related as $1.7$ before the ``knee'' and $2.0$ after the knee. Drift processes due to the galactic magnetic field can be responsible for this small discrepancy, as well as near galactic sources, such as a supernova remnant,
could be responsible for the ``knee'' in the energy spectrum. In addition, the index as $3/2$ is also close to the integral index of the power energy spectrum of extra-galactic origin related as $1.7$ in the energy region above $10^{18}$ eV. 

If the extra-galactic cosmic rays are cosmological, we have a direct information on the energy distribution of the Universe. The index of its energy spectrum as 1.7 is plausible for the equipartition of energy into the three modes under consideration, and apparently, there is not room for other energy modes like the dark energy. The situation is dramatic, because 73\% of an ``unknown'' dark energy in the Universe is required in the Big Bang cosmology.

\section{The ultra high energy cosmic ray and the CMB}

So far, cosmic rays with energies above the GZK have observed practically in all extensive air shower experiments, and the record goes to the AGASA experiments with 11 events above the GZK cutoff.
Some characteristics of UHECR can be obtained from a global study of the data available for instance  in the World Data Center C2 for Cosmic Rays, Institute of Physical and Chemical Research Itabashi, Tokyo, Japan or in the Yakutsk EAS array catalog \cite{yakutsk}.
The main characteristic of the UHECR is the isotropy observed in their arrival direction distribution as shown in Fig.1. Strictly speaking, the UHECR distribution is almost isotropic, because
some small anisotropies on the scale of few degrees has been found in the AGASA data \cite{takeda99} in the form of doublets and triplets of events.

The energy distribution of the UHECR in the three largest extensive air shower experiments is shown in Fig.2.
The absence of the GZK cutoff specially in the AGASA experiment is evident. No acceleration mechanism is known to such high energies within our local cluster, and they cannot be cosmological because of this GKZ cut-off.

Here we argue that the hypothesis of a local origin of the CMB can explain the absence of the GZK cutoff in the UHECR energy spectrum. In this scenario, the CMB radiation originates relatively close to us, the CMB is produced by early generation of stars in the 4He synthesis \cite{cuybert04}, because the amount of energy released in producing the observed amount of 4He
is the same as the amount of energy in the CMB. The CMB radiation is thermalized and isotropized by a thicket of dense,
magnetically confined plasma filaments in the intergalactic medium of our local super cluster of galaxies \cite{lerner88}. The average mass and size of a cluster can be obtained from the dispersion velocity of galaxies in the cluster. The ESO survey \cite{ramella02} supply
\begin{equation}
\bar{M}_{group}\approx 1.15 \times 10^{14} M_{\odot},
\end{equation}
where $M_{\odot}$ is the solar mass.  The size of cluster is obtained from the equation
\begin{equation}
\frac{GM}{R}=2\sigma^2,
\end{equation}
where $\sigma\approx 270 km s^{-1}$ is the dispersion velocity given by the ESO survey . This value provides a cutoff radius $\bar{R}\approx 3$ Mpc.  Thus, the CMB is concentrated basically inside a spheres of $\sim 3 Mpc$ of radius. This value is smaller than the attenuation length of ultra high energy protons $\sim 10^{20}$ eV in the CMB and which is estimated as $20$ Mpc.

In the past years, new data on UHECR were available from the HiRes fluorescence detector
\cite{abu-zayyad}, an upgrade of the Fly's Eyes detector at Utah, USA. So far, only one event with energy above $10^{20}eV.$ has been observed. However, it is observed a systematic behavior in the energy spectrum above $10^{19}eV$ that is in agreement with the existence of a GZK cutoff when an isotropic distribution of sources is assumed.
In addition, in recent paper of the HiRes Group present at 30th ICRC (Merida, Mexico, 2007) \cite{abbasi07}, it is claimed a definitive observation of the GZK cutoff. The result is extremely dependent of Monte Carlo calculations , set up starting from extrapolations of models and data of particle's accelerators to higher energies.
The poor number the sampled events with energies above $10^{20}eV$ still doesn't allow us to define if the cutoff really exists. Even so, the data of the AGASA is more robust, at least statistically speaking.
We are waiting for the next round of experiments such as AUGER (air shower plus fluorescence), EUSO, and OWL (both fluorescent detector at spacecraft) that can increase the statistics by a factor of 2 orders of magnitude.

\section{Conclusions}

In 2003, results became available from the Wilkinson Microwave Probe (WMAP) satellite, which showed the CMB radiation in more details than previously. The new results were interpreted as a complete confirmation of the Big Bang cosmology.
The observed smoothness of the CMB in WMAP data was described as a success of the  inflation hypothesis. The first stage of the nascent universe would have passed through a phase of exponential expansion that was driven by a negative pressure vacuum energy density. However, subsequent analysis has shown some inconsistences with the hypothesis of a cosmological origin of the CMB. The main ones are the absence of gravitational lensing effect in the WMAP data, as well as the non random distribution of the observed cold spots. It is possible to see several alignments specially with the local cluster. The absence of gravitational lensing in the CMB and the alignments of the CMB strongly constrain the assumption that the CMB is of cosmological origin. The CMB would be produced by early generation of stars in the 4He synthesis \cite{cuybert04}.

In addition, the cluster shadow effect on the CMB has shown that the CMB from behind the cluster is slightly shadowed by hot electrons in the cluster. There is in fact no strong evidence in the WMAP database for the Sunyaev-Zel'dovich effect. There is no strong evidence for a emission origin of the CMB at locations beyond the cluster.
The average size of a cluster can be obtained from the dispersion velocity of galaxies in the cluster. The ESO survey supply an average value of $\bar{R}_{cluster}\sim 3$ Mpc. Consequently we argue that the CMB is confined basically inside a sphere with the same size of the cluster. The extension of $\sim 3$ Mpc of a cluster is smaller than than the attenuation length of ultra high energy protons $\sim 10^{20}$ eV in the CMB which is estimated as $20$ Mpc. Under this hypothesis, the universe would be almost transparent for the UHECR.
  
The analysis on the basis of the cosmic energy spectrum within the energy equipartition theorem strongly constrain the dark matter and dark energy  assumption, essential for the Big bang cosmology. In addition,
the hypothesis of a local origin for the CMB linked only to the local super-cluster
takes into account all global characteristics of the UHECR, such as the isotropic distribution of their arrival directions and a possible absence of the GZK cutoff in the energy spectrum. But important questions such as acceleration mechanism of the UHECR particles remain open, and require continuous observations and further investigations. The ongoing Auger experiment will deliver much better statistics in the coming years,
and the Planck satellite that probably will be launched in 2008 will also deliver much better information on the CMB.
Only like this, we will be able to confirm or refute the hypothesis of a local origin of CMB 
as responsible of the global characteristics observed in the UHECR.


\begin{figure}[th]
\includegraphics[clip,width=0.6
\textwidth,height=0.6\textheight,angle=0.] {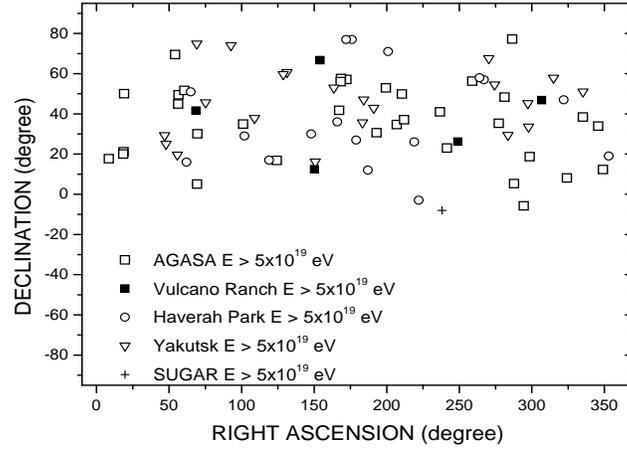}
\vspace*{-7.0cm}
\caption{Arrival direction distribution of cosmic rays with energy$\geq 5\times 10^{19}$}%
\end{figure}

\begin{figure}[th]
\vspace*{-1.0cm}
\includegraphics[clip,width=0.6
\textwidth,height=0.6\textheight,angle=0.] {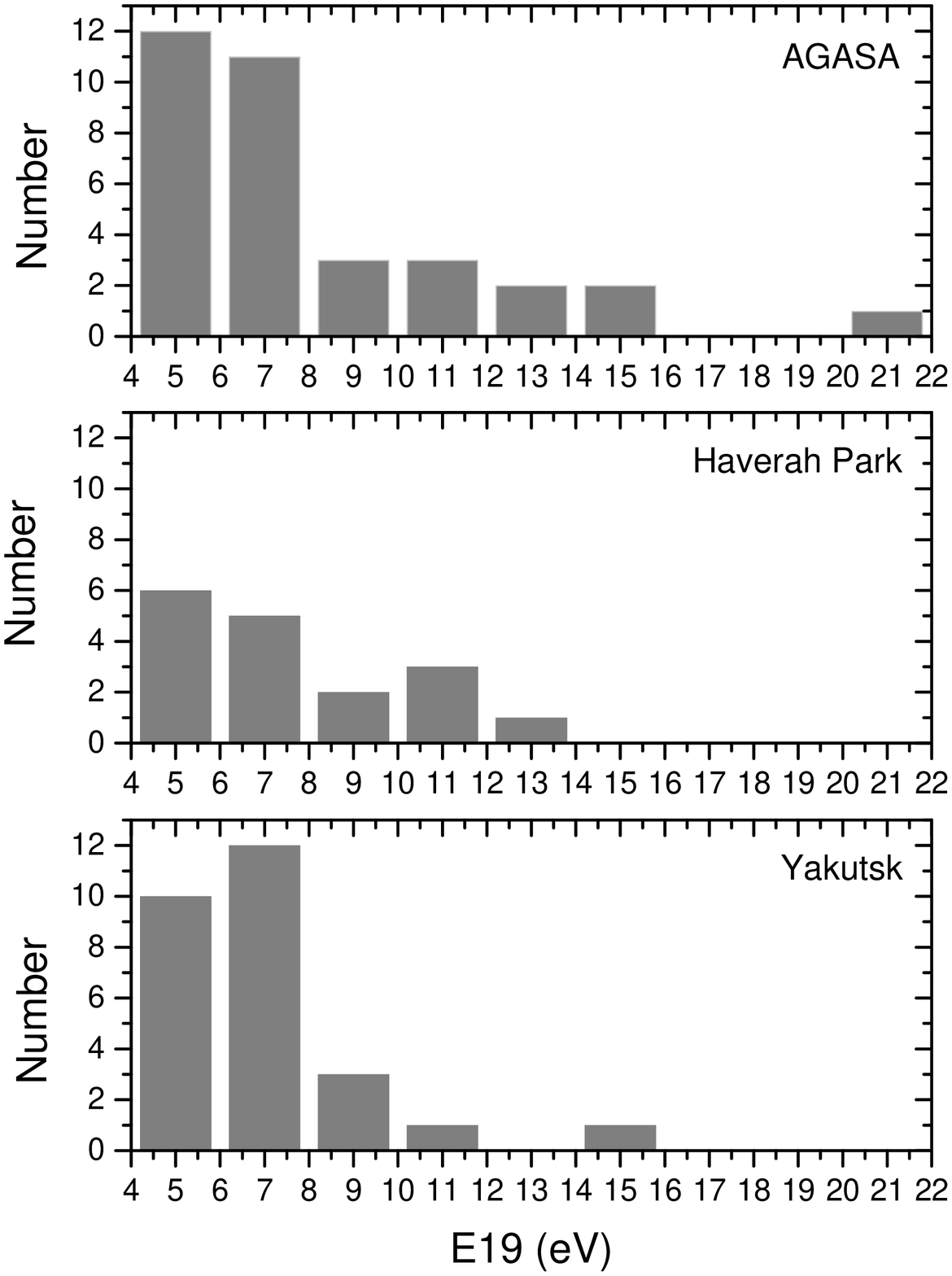}
\caption{Energy distribution of cosmic rays with energy$\geq 5\times 10^{19}$.}%
\end{figure}

\end{document}